\documentclass{emulateapj}

\slugcomment{draft version \today, accepted for publication in the Astrophysical Journal}

\shortauthors{Ito et al.}

\begin{document}

\title{
  EVOLUTION OF NON-THERMAL EMISSION FROM SHELL ASSOCIATED WITH AGN JETS}

\author{Hirotaka Ito\altaffilmark {1}, Motoki Kino\altaffilmark{2},
    Nozomu Kawakatu\altaffilmark{3},
    and
    Shoichi Yamada\altaffilmark{4,5} }


\altaffiltext{1}{
Department of Aerospace Engineering, Tohoku University, 6-6-01 Aramaki-Aza-Aoba, Aoba-ku, Sendai, 980-8579, Japan}
\email{hito@rhd.mech.tohoku.ac.jp}
\altaffiltext{2}{National Astronomical Observatory of Japan, 2-21-1
Osawa, Mitaka, Tokyo 181-8588, Japan}
\altaffiltext{3}{Graduate School of Pure and Applied Sciences, University of Tsukuba, 1-1-1 Tennodai, Tsukuba 305-8571 kawakatu@ccs.tsukuba.ac.jp}
\altaffiltext{4}{Science and Engineering, Waseda University, 3-4-1 Okubo,
Shinjuku, Tokyo 169-8555, Japan}
\altaffiltext{5}{Advanced Research Institute for Science \&
Engineering, Waseda University, 3-4-1 Okubo,
Shinjuku, Tokyo 169-8555, Japan}







\begin{abstract}

 We explore the evolution of the emissions by accelerated electrons in
 shocked shells driven by jets in active galactic nuclei (AGNs). 
 Focusing on powerful sources which host luminous quasars, 
 we evaluated the broadband emission spectra by properly taking into
 account adiabatic and radiative cooling effects on the
 electron distribution.  The synchrotron radiation 
 and inverse Compton (IC) scattering of various  photons
 that are mainly produced in the accretion disc and dusty torus
 are considered as radiation
processes.
 We show that
 the resultant radiation is dominated by the IC emission for compact 
 sources ($\lesssim 10{\rm kpc}$), whereas the synchrotron radiation 
is more important for larger sources.
 We also compare the shell
 emissions with those expected from the lobe
 under the assumption  that a fractions of
 the energy deposited in the shell and lobe
 carried by the non-thermal electrons are $\epsilon_e \sim 0.01$ and $\epsilon_{e, {\rm lobe}} \sim 1$, respectively. 
 Then,
 we find that the shell emissions
 are brighter than the lobe ones at 
 infra-red and optical bands 
 when the source size is $\gtrsim 10{\rm kpc}$, and
 the IC emissions from the shell
 at $\gtrsim 10~{\rm GeV}$ 
 can be observed with the absence of contamination from the lobe 
 irrespective of the source size.   
 In particular, it is predicted that,
 for most powerful nearby sources
 ($L_{\rm j} \sim 10^{47}~{\rm ergs~s^{-1}}$), $\sim {\rm TeV}$
  gamma-rays produced via
 the IC emissions can be detected by the modern Cherenkov telescopes
 such as MAGIC, HESS and VERITAS.

\end{abstract}


\section{INTRODUCTION}

 It is well established that radio-loud active galactic nuclei
 (AGNs) are accompanied by relativistic jets
 \citep[e.g.,][for review]{BBR84}.
 These jets dissipate their kinetic energy via
 interactions with surrounding  interstellar medium (ISM) or
 intracluster medium (ICM), 
 and inflate a bubble composed of decelerated jet matter,
 which is often referred to as cocoon. 
 Initially, the cocoon is highly overpressured against 
 the ambient ISM/ICM  \citep{BC89} and a strong  shock is driven 
into the ambient matter. Then a thin shell is formed around the cocoon 
by the compressed ambient medium.
The  thin shell structure persists until the cocoon pressure decreases and
the pressure equilibrium is eventually achieved \citep{RHB01}.

 While large number of  radio observations
 identified  cocoons with the 
 extended radio lobe, 
 no clear evidence of radio emissions
 is found for the shocked shells  \citep[e.g.,][]{CPD88}.
 Due to the lack of detections,
 in the previous studies on the extragalactic radio sources,
 it is usually assumed that non-thermal emissions
 are dominated by or
 originated only in the cocoon \citep[e.g.,][]{SBM08}. 
 However,
 since strong shocks are driven into tenuous ambient gas
 with a high Mach number,
 the shocked shells are expected to
 offer site of particle acceleration
 as 
 in the shocks of supernova remnants (SNRs) 
 and therefore give rise
 non-thermal emissions
 \citep{FKY07, B08}.
 Hence, although the 
 observations at radio 
 seems to be unsuccessful,
 non-thermal emissions 
 from the shell may
 be accessible  at higher frequencies.
 In fact,
 while not detected in radio,
 recent deep X-ray observation 
 have reported 
 the presence of  non-thermal emissions  from the shell
 associated with the radio galaxy Centaurus A
 \citep{CKH09}.
%
%
%

%




%

%

 Theoretically, 
 emissions by the accelerated particles
 residing in shocked shells 
 have been studied by \citet{FKY07}.
 However,  they paid attention only to the
 the extended sources of $\sim 100~{\rm kpc}$
 and the inverse Compton (IC) scattering of external photons
 was not included in the radiative processes.
 The cooling effects on the energy distribution of non-thermal electrons
 were not considered, either.
%
 Motivated by these backgrounds, we  explore in this paper the 
 temporal 
 evolution of the non-thermal emissions by the accelerated electrons in 
the shocked shells, properly taking into account the Comptonization of photons
 of various origins as well as the
 cooling effects on the electron distribution. 
 Focusing on the powerful sources
 which host luminous quasar in its core,
 we show that the shell
 can produce prominent emissions ranging from 
 radio up to $\sim 10~{\rm TeV}$ gamma-ray and
 discuss the possibility for the detection.

The paper is organized as follows.
 In \S\ref{model}, we introduce 
 the dynamical model, which describes
 the evolution of the shell expansion, and
 explain how the energy distribution of electrons
 residing in the shell
 and
 the spectra of the radiations they produce are evaluated based on 
 the dynamical model. 
 The obtained results are presented in \S\ref{results}.
 In \S\ref{comp}, we
 compare the emissions from the shell with those from the lobe 
 and discuss its detectability.
 We close the paper with the summary in \S\ref{summ}.

\section{MODEL FOR  EMISSIONS FROM THE SHELL}
\label{model}

\subsection{Dynamics}
\label{dynamics}

\begin{figure}[h]
\begin{center} 
\includegraphics[width=8.4cm]{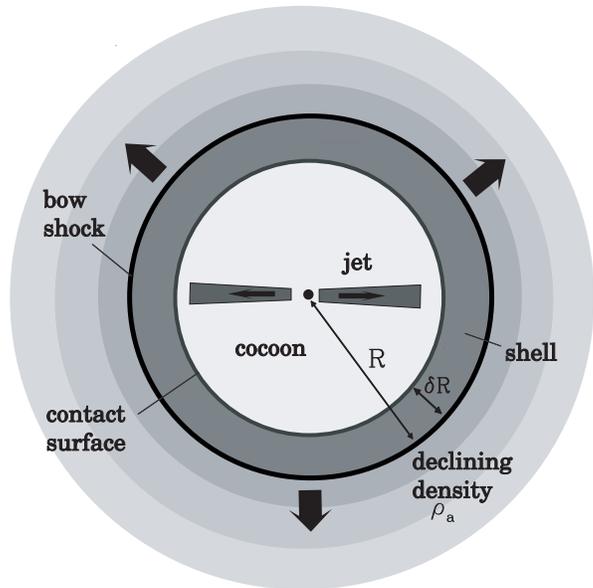}
\caption
{The schematic picture of the model employed in this study   
 to approximate the expansion of the cocoon
 and shell produced by the jet pushing into the 
 ambient medium from AGN.
}
\label{cocoon}
\end{center}
\end{figure}

  In this section, we give the model to approximately describe the dynamics of the 
expanding cocoon and shell and to provide a basis, on which the energy distribution of 
accelerated electrons and the radiation spectra are estimated.  
  The schematic picture of the model  
  is illustrated in Fig. \ref{cocoon}.
  For simplicity we neglect the elongation of cocoon and shell in the jet direction 
  and assume that they are spherical.
  We also assume that the shell width, $\delta R$, is thin compared with 
  the size of the whole system, $R$, and 
 ignore the difference between the radii of the bow shock
 and the contact surface.
  This assumption is valid 
 as long as the expansion velocity has
 a high Mach number 
 \citep[see e.g.,][]{OM88}.
 We further assume that the kinetic power of jet, $L_{\rm j}$, is constant in time on 
the time scale of relevance in this study.

Under the above assumptions, the dynamics of the expanding 
cocoon and shell can be approximately described by
the model of  stellar wind bubbles \citep{CMW75}.
The basic equations are 
 the followings: 
(\ref{mom}) the equation for the momentum of the swept-up matter in the shell and
(\ref{energy}) the equation for the energy in the cocoon.
They are expressed, respectively, as
\begin{eqnarray}
\frac{d}{dt}\left(M_{\rm s}(t) \dot{R}(t) \right)=
 4 \pi R(t)^2 P_{\rm c}(t),
\label{mom}
\end{eqnarray}
\begin{eqnarray}
\frac{d}{dt}
\left( \frac{P_{\rm c}(t)V_{\rm c}(t)}{\hat{\gamma_{\rm c}}-1} \right)
  + P_c(t)\frac{dV_{\rm c}(t)}{dt}
=  L_{\rm j}    ,
\label{energy}
\end{eqnarray}
where
$\dot{R} = dR/dt$,
$V_{\rm c} = 4 \pi R^3 /3 $, and 
$P_{\rm c}$
 are 
 the expansion velocity of the shell, the volume and  
 pressure of the cocoon, respectively,
 and $\hat{\gamma}_{\rm c}$ is
 the specific heat ratio for the plasma inside the cocoon.
The swept-up mass in the shell is defined as
 $M_{\rm s}=\int_{0}^{R}4 \pi r^2 \rho_{\rm a}(r) dr$ 
 with the mass density of the ambient medium, $\rho_{\rm a}$.

%

 In this study, $\rho_{\rm a}$  is assumed to be given by
 $\rho_{\rm a}(r) ={\rho_0}(r/{ r_0})^{- \alpha}$ with 
 ${r_0}$ and ${\rho_0}$ being the reference position and
 the mass density there, respectively.
 Then Eqs. (\ref{mom}) and (\ref{energy}) can be solved analytically and 
 the solution is given as 
\begin{eqnarray}
\label{R}
R(t)=C_{R}~
            { r_0}^{\alpha/(\alpha-5)}
             \left(\frac{ L_{\rm j}}{\rho_0} \right)^{1/(5-\alpha)}
          t^{3/(5-\alpha)} ,
\end{eqnarray}
 where $C_{R}$ is a function of $\alpha$ and 
 $\hat{\gamma_{\rm c}}$ given as
\begin{eqnarray}
C_{R} = 
\left[ \frac{(3-\alpha)(5-\alpha)^3(\hat{\gamma_{\rm c}}-1)}
{4\pi \{ 2\alpha^2 + (1-18\hat{\gamma_{\rm c}})\alpha + 63\hat{\gamma_{\rm
c}}-28 \}  }
 \right]^{1/(5-\alpha)}.
\nonumber
\end{eqnarray}
%
 
Employing the non-relativistic Rankine-Hugoniot condition in the strong shock limit~\citep{LL59}, 
which is justified by the assumption that the shell expands with a high Mach number, 
 we obtain from Eq. (\ref{R}) the density, $\rho_{\rm s}$, and pressure, $P_{\rm s}$, in the shell, 
 which are assumed to be uniform, as follows:
\begin{eqnarray}\label{rhos}
\nonumber  \rho_{\rm s}(t)
             =  \frac{{\hat \gamma}_{\rm a} + 1}{{\hat \gamma}_{\rm a} - 1}
                  {\rho_0}
                  \left(\frac{R(t)}{r_0} \right)^{-\alpha}, \\ 
%
P_{\rm s}(t) = \frac{2}{\hat{\gamma}_{\rm a} + 1}
             {\rho_0}
             \left(\frac{R(t)}{ r_0} \right)^{-\alpha} 
             {\dot R}(t)^2  ,
\end{eqnarray}
  where ${\hat \gamma}_{\rm a}$ is the specific heat ratio
   for the ambient medium.
  By equating $M_{\rm s}$ with the  mass in the shell given by 
 $V_{\rm s} \rho_{\rm s}$, where $V_{\rm s} = 4 \pi R^2 \delta R$ is the volume of the shell,
 the shell width is obtained as
$\delta R = ({\hat \gamma}_{\rm a} - 1)
  [({\hat \gamma}_{\rm a} + 1)(3-\alpha)]^{-1} R $ . 
 It is found that the ratio of $\delta R$ to $R$ does not depend on time and
 the thin-shell approximation is reasonably good for the typical values, 
${\hat \gamma}_{\rm a} = 5/3$ and $0 \leq \alpha \leq 2$.
 

  The total internal energy stored in the shell,
  $E_{\rm s} = P_{\rm s} V_{\rm s} / ({\hat{\gamma}_{\rm a}} - 1)$, is
one of the most important quantities, since it determines the energy budget 
for the radiation. From Eqs. (\ref{R}) and (\ref{rhos}), $E_{\rm s}$
  can be expressed as
\begin{eqnarray}
  E_{\rm s} = f L_{\rm j} t,
\label{Es}
\end{eqnarray} 
where $f$ is a fraction of the total energy released by the 
  jet ($L_{\rm j} t$), which is converted to the internal energy
  of the shell, and is given by 
\begin{eqnarray}
   f = \frac{18 ({\hat{\gamma}_{\rm c}} - 1) (5-\alpha)}
           {({\hat{\gamma}_{\rm a}} + 1)^2
             [2\alpha^2 + (1-18{\hat{\gamma}_{\rm c}}) \alpha +
              63 {\hat{\gamma}_{\rm c}} - 28]} ,
\label{factor}
\end{eqnarray} 
  which turns out to be time-independent. For typical numbers, ${\hat \gamma}_{\rm c} = 4/3$, 
  ${\hat \gamma}_{\rm a} = 5/3$, $0 \leq \alpha \leq 2$, $f$ depends on $\alpha$
 only weakly and $f \sim 0.1$.

\subsection{Energy Distribution of Electrons}
\label{ele}

 Assuming that a fraction, $\epsilon_e$, of the energy deposited in the shell, $E_{\rm s}$, 
 goes to the non-thermal electrons, we evaluate their energy 
 distribution.
%
%
 We solve the kinetic equation  governing the temporal evolution of 
 the energy distribution of electrons, $N(\gamma_e, t)$,
  which is given as
\begin{eqnarray}
 \label{kinetic}
  \frac{\partial N(\gamma_e, t)}{\partial t} =
 \frac{\partial}{\partial \gamma_e}
  [ \dot{\gamma}_{\rm cool}(\gamma_e) N(\gamma_e, t) ]
 + Q(\gamma_e) ,
\end{eqnarray}
 where $\gamma_e$, $\dot{\gamma}_{\rm cool}(\gamma_e) = - d\gamma_e/dt$,
 and $Q(\gamma_e)$ are the Lorentz factor, the cooling rate via adiabatic expansions and
 radiative losses, and the injection rate of non-thermal electrons, respectively.
 The latter two, that is, the injection rate
 $Q(\gamma_e)$ and the cooling rate
 $\dot{\gamma}_{\rm cool}$,
 which will be described in detail below,
 are evaluated based on the dynamical model described in the previous section.



 Following the theory of the diffusive shock acceleration (DSA) \citep{B78, D83, BE87},
 we assume that the non-thermal electrons are injected into
 the post-shock region with a power-law energy distribution of the form
\begin{eqnarray}
  \label{Qinject}
Q(\gamma_e) =
                K \gamma_e^{-p}
         ~~{\rm for}~~  \gamma_{\rm min}\leq \gamma_e
                       \leq \gamma_{\rm max},
\end{eqnarray}
 where
 $\gamma_{\rm min}$ and $\gamma_{\rm max}$ correspond
 to the minimum and maximum Lorentz factors, respectively. 
 In this study we set $\gamma_{\rm min} = 1$ and the power-law index,
 $p$, is fixed to $2$, an appropriate value for the linear diffusive shock acceleration 
 in the strong shock limit.
 The maximum Lorentz factor is obtained by equating the 
 the cooling rate, $\dot{\gamma}_{\rm cool}$, to the  acceleration rate given by 
\begin{eqnarray}
  \label{accel}
  \dot{\gamma}_{\rm accel} = \frac{3 e B \dot{R}^2}{20 \xi  m_e c^3} , 
\end{eqnarray} 
 where $B$ is the magnetic field strength in the shell.
 The so-called ``gyro-factor'', $\xi$, can be identified with  the ratio of the energy
 in ordered magnetic fields to that in turbulent ones ($\xi = 1$ is the Bohm limit).
 The normalisation factor, $K$, in the injection rate is determined from the assumption 
 that a fraction, $\epsilon_e$, of the shock-dissipated energy is carried by the 
 non-thermal electrons: 
 $\int^{\gamma_{\rm max}}_{\gamma_{\rm min}} 
   (\gamma_e - 1) m_e c^2 Q(\gamma_e) d\gamma_e=
 \epsilon_e dE_{\rm s}/dt = f \epsilon_e L_{\rm j}$.
 A rough estimation of the normalization factor is 
 obtained as
 $K \sim 0.1 \epsilon_e L_{\rm j} / [m_e c^2{\rm ln}({\gamma_{\rm max}})]$,
 where we used $f \sim 0.1$ (\S\ref{dynamics}).
 It is noted that since 
 the factor $K$ is proportionate to $\epsilon_e$ and $L_{\rm j}$, 
 the resultant luminosity of non-thermal emissions also scales in the same 
 manner with these quantities.

 In the cooling rate, $\dot{\gamma}_{\rm cool}$, both radiative and adiabatic 
 losses are taken into account.
 As for the former, 
 the synchrotron radiation and the IC scattering off photons of various origins.
%
%
%
 The cooling rate for the adiabatic dynamical expansion is expressed for electrons with 
a Lorentz factor $\gamma_e$ as
\begin{eqnarray}
 \label{adiabatic}
  \dot{\gamma}_{\rm ad} = \frac{1}{3}\frac{\dot{R}}{R}\gamma_e .
\end{eqnarray}
%
 The rates of the radiative coolings via the synchrotron radiation and IC emissions are given, 
 respectively, as
\begin{eqnarray}
 \label{synchro}
  \dot{\gamma}_{\rm syn} = \frac{4\sigma_T  U_B}{3 m_e c}
   \gamma_e^2 ,
\end{eqnarray} 
 and
\begin{eqnarray}
 \label{IC}
  \dot{\gamma}_{\rm IC} =  \frac{4\sigma_T  U_{\rm ph}}{3 m_e c}
 \gamma_e^2 F_{\rm KN}(\gamma_e),
\end{eqnarray} 
 where $\sigma_T$, $c$ and $m_e$ are
 the cross section of Thomson scattering, the speed
 of light and the electron mass, respectively.
 The energy densities of magnetic fields and photons in the 
 shell are denoted by $U_B = B^2 / 8 \pi$ and $U_{\rm ph}$, respectively.
 The function $F_{\rm KN}(\gamma_e)$ encodes both the distributions of
 seed photons and the Klein-Nishina (KN) cross section for the Compton 
 scattering and reduces to unity in the Thompson limit. Note, however, 
 we do not employ this limit but calculate $F_{\rm KN}(\gamma_e)$
 from the differential cross-section for IC scattering given 
 in \citet{BG70}.


 The typical magnetic field strengths in elliptical galaxies and clusters of
 galaxies are a few ${\rm \mu G}$ \citep[e.g.,][]{MS96, VMM01, CKB01, CT02, SCK05}.
Assuming that as it passes through a shock wave, the magnetic field is adiabatically 
compressed by a factor of $1 \sim 4$ depending on the obliqueness of the
 shock to the magnetic field lines, we choose $B = 10{\mu {\rm G}}$ as a fiducial value
 for the magnetic field strength in the shell when evaluating 
the acceleration rate, $\dot{\gamma}_{\rm accel}$,  and
 synchrotron cooling rate, $\dot{\gamma}_{\rm syn}$.
%
%
%
 The corresponding energy density of magnetic field is given by
\begin{eqnarray}
\label{UB}
 U_{\rm B} \approx 4.0 \times 10^{-12}B_{-5}^2~{\rm
 erg~cm^{-3}},
\end{eqnarray}
 where $B_{-5} = B / 10{\rm \mu G}$.

 In evaluating  the cooling rate for IC scattering, $\dot{\gamma}_{\rm IC}$,
 we take into account various seed photons of relevance in this context. 
 In their paper, \citet{SBM08} explored  high energy 
 emissions by relativistic electrons in the radio lobes 
 through the IC scatterings of various photons. Among them are UV emissions 
from the accretion disc, IR emissions from the dusty torus, stellar emissions 
in NIR from the host galaxy and synchrotron emissions from the radio lobe.
In addition to these emissions, we also consider CMB  as seed photons in the 
present study.
Following \citet{SBM08}, we assume that the photons from the disc,  torus,  host galaxy,
 and CMB are monochromatic and have the following single frequencies:
 $\nu_{\rm UV} = 2.4 \times 10^{15}~{\rm Hz}$, 
 $\nu_{\rm IR} = 1.0 \times 10^{13}~{\rm Hz}$, 
 $\nu_{\rm NIR} = 1.0 \times 10^{14}~{\rm Hz}$, 
 and
$\nu_{\rm CMB} = 1.6 \times 10^{11}~{\rm Hz}$.
The photons from the radio lobe are assumed to have a continuous spectrum 
given by $L_{\rm \nu, lobe} \propto \nu^{-0.75}$.

From the luminosity, $L_{\rm UV}$, of the UV emissions by the accretion disc,
 the energy density of these photons in the shell is given approximately as
\begin{eqnarray}  
 \label{UUV}
  \nonumber
  U_{\rm UV} & = & \frac{L_{\rm UV}}{4 \pi R^2 c}  \\ & \approx &
     3.0 \times 10^{-9}L_{\rm UV, 46}
                      R_1^{-2}~{\rm erg~cm^{-3}} ,
\end{eqnarray}
 where $L_{\rm UV,46} = L_{\rm UV}/10^{46}{\rm erg~s^{-1}}$,
 and $R_1 = R/1{\rm kpc}$.
 As is assumed in \citet{SBM08}, we take 
 $L_{\rm UV} = 10^{46}~{\rm ergs~s^{-1}}$
 for sources with jet kinetic power of $L_{\rm j} > 10^{45}~{\rm ergs~s^{-1}}$,
 and $L_{\rm UV} = 10^{45}~{\rm ergs~s^{-1}}$ for
 $L_{\rm j} \leq 10^{45}~{\rm ergs~s^{-1}}$
 as fiducial values.
 The adopted values are appropriate for sources hosting a luminous quasar.
 According to the broadband observations of quasars 
 \citep[e.g.,][]{E94, J06}, the IR and UV emissions are approximately comparable although 
 there are some variations in the relative strengths from source to source.
 We hence assume that  the luminosity of the IR emissions from the torus, $L_{\rm IR}$,
 is equal to that of the UV emissions from the disc, $L_{\rm UV}$, and
 the energy density of IR photons in the former case is estimated as
\begin{eqnarray}  
 \label{UIR}
  U_{\rm IR} = \frac{L_{\rm IR}}{4 \pi R^2 c} \approx
     3.0 \times 10^{-9} L_{\rm IR, 46}
                      R_1^{-2}~{\rm erg~cm^{-3}} ,
\end{eqnarray}
 where
 $L_{\rm IR,46} = L_{\rm IR}/10^{46}~{\rm erg~s^{-1}}$.
 In evaluating the energy density of photons from
 the host galaxy, we assume that most of the photons are emitted by stars in the  
 the core region with the radius of $\sim 1~{\rm kpc}$ \citep{RPC05}.
 Considering only the sources with $R \gtrsim ~{\rm kpc}$, we can estimate
 the energy density of the optical photons from the luminosity, $L_{\rm NIR}$,
 as 
\begin{eqnarray}  
 \label{UV}
  \nonumber
  U_{\rm NIR} & = & \frac{L_{\rm V}}{4 \pi R^2 c} \\
  & \approx  &
3.0 \times 10^{-10} L_{\rm NIR, 45}
                      R_{1}^{-2}~{\rm
		      erg~cm^{-3}},
\end{eqnarray}
 where 
 $L_{\rm NIR, 45} = L_{\rm NIR}/10^{45}{\rm ergs~s^{-1}}$.
 The energy density of CMB photons is  given by
\begin{eqnarray}  
 \label{UCMB}
  U_{\rm CMB} \approx 4.2 \times 10^{-13} 
 ~{\rm  erg~cm^{-3}} .
\end{eqnarray}
 The redshift is ignored for simplicity. Finally, the energy density of photons 
emitted from the lobe is obtained by assuming that
 a fraction $\eta$ of the jet power is radiated as radio emissions from the lobe 
 (i.e., $\int L_{\rm \nu, lobe}d\nu = \eta L_{\rm j}$).
 Assuming $\eta = 10^{-2}$ as a fiducial case, 
 it is given as
\begin{eqnarray}  
  \label{Ulobe}
  \nonumber
  U_{\rm lobe} & = & \frac{\int L_{\rm \nu, lobe} d\nu}{4 \pi R^2 c} \\
  & \approx &
     3.0 \times 10^{-12}
{\eta_{-2}} L_{45} R_1^{-2}~{\rm  erg~cm^{-3}} ,
\end{eqnarray}
 where 
 $\eta_{-2} = \eta / 10^{-2}$ and
 $L_{45} = L_{\rm j}/10^{45}~{\rm ergs~s^{-1}}$.

The cooling rate for the IC scattering is
obtained from Eq. (\ref{IC}) by plugging in the energy densities
 $U_{\rm ph}$ obtained above and calculating $F_{\rm KN}(\gamma_e)$ for 
the assumed photon distributions.
 Each contribution from the disc, torus, host galaxy, CMB, and
 lobe is denoted as
 $\dot{\gamma}_{\rm IC,UV}$, 
 $\dot{\gamma}_{\rm IC,IR}$,
 $\dot{\gamma}_{\rm IC,NIR}$,
 $\dot{\gamma}_{\rm IC,CMB}$, 
 and
 $\dot{\gamma}_{\rm IC,lobe}$.
 The total cooling rate $\dot{\gamma}_{\rm cool}$
 is the sum of the rates for the adiabatic, synchrotron and  IC losses.
 ($\dot{\gamma}_{\rm cool} =
    \dot{\gamma}_{\rm ad} +
    \dot{\gamma}_{\rm syn} +
    \dot{\gamma}_{\rm IC,UV} +
    \dot{\gamma}_{\rm IC,NIR} +
    \dot{\gamma}_{\rm IC,CMB} +
    \dot{\gamma}_{\rm IC,lobe} 
$).
Note that the synchrotron self-Compton is ignored, since its
 effect is negligible in any case considered in the present study. 
%

%
%

%

%
 Now that the injection rate, $Q(\gamma_e)$, and the cooling rate, $\dot{\gamma}_{\rm cool}$,
 have been evaluated, the energy distribution of non-thermal electrons, $N(\gamma_e, t)$,
 is obtained by putting these quantities in Eq. (\ref{kinetic}).
 Although $Q(\gamma_e)$ is time-dependent through $\gamma_{\rm max}$ and so is $\dot{\gamma}_{\rm cool}$ 
because of the power-dependence of the expansion rate (see Eq.~(\ref{R})), we ignore  
these variations over the dynamical time scale $\sim t$. Then, employing the instantaneous 
 values of $\dot{\gamma}_{\rm cool}(\gamma_e)$
 and $Q(\gamma_e)$ evaluated at each time and fixing them, we can solve Eq. (\ref{kinetic}) by the 
 Laplace transforms \citep{M80, MHK07} and the solution can be written as
\begin{eqnarray} 
 \label{N}
 N(\gamma_e, t) = \frac{1}{\dot{\gamma}_{\rm cool}(\gamma_e)}
  \int^{\gamma_{\rm u}}_{\gamma_e} d\gamma_e' Q(\gamma_e'),
\end{eqnarray}
 where the upper limit of the integral is given by
\begin{eqnarray}
 \nonumber
  \gamma_{\rm u} = 
  \left\{ \begin{array}{ll}
      \gamma_*(\gamma_e) &~~
         {\rm for}~~ \gamma_{\rm min} \leq \gamma_e \leq  
                  \gamma_{\rm br} ,  \\
       \gamma_{\rm max} &~~
         {\rm for}~~ \gamma_{\rm br} < \gamma_e \leq  
                  \gamma_{\rm max} .  \\
             \end{array} \right.
\end{eqnarray}
 Here $\gamma_*$ and $\gamma_{\rm br}$ 
 are determined by the relation 
 $\tau(\gamma_*, \gamma_e) =
 \tau(\gamma_{\rm max}, \gamma_{\rm br}) = t$,
 where the function
 $\tau(\gamma', \gamma) = \int^{\gamma'}_{\gamma}  d\gamma''
  / \dot{\gamma}_{\rm cool}(\gamma'')   $
 gives the time it takes an electron to cool from
 $\gamma'$ to $\gamma$ in the Lorentz factor.
 Note that $\gamma_{\rm br}$ corresponds to the Lorentz factor,
 where the energy distribution of non-thermal electrons shows a break
 owing to the radiative coolings  (see \S\ref{ele2} for detail).
%

\subsection{Radiation spectra}
\label{spectra}


 From the energy distribution of non-thermal electrons just obtained,
the spectrum of the synchrotron radiation is calculated as
\begin{eqnarray}
  L_{\rm \nu, syn} = \int^{\gamma_{\rm max}}_{\gamma_{\rm min}}
   P_{\rm syn}(\nu, \gamma_e) N(\gamma_e) d \gamma_e .
 \label{Lsyn}
\end{eqnarray}
 Here, $P_{\rm syn}(\nu, \gamma_e)$ is the pitch-angle-averaged
 power spectrum for a single electron given by
\begin{eqnarray}
\nonumber
  P_{\rm syn}(\nu, \gamma_e) =
 \frac{\sqrt{3} e^3 B}{2 m_e c^2}
 \int^{\pi}_{0} {\rm sin}^2\theta F\left(\frac{\nu}{\nu_{\rm c}} \right)
 d\theta,
\end{eqnarray}
  where $\theta$ is the pitch angle and $\nu_{\rm c} = 3 e B \gamma_e^2 {\rm sin}\theta / 4 \pi m_e c$
 is the characteristic frequency of the emitted photons.
 The function $F(x)$ is defined by
 $F(x) = x \int^{\infty}_x K_{5/3}(y)dy$, where $K_{5/3}(y)$ is the 
 modified Bessel function of the $5/3$ order \citep{RL79}.
 Note that the synchrotron self-absorption is ignored,
 since it is important only at low frequencies below $10^{7}{\rm Hz}$
 in the current context.

 In the same way, the spectrum of the IC scattering is obtained as
\begin{eqnarray}
 \label{LIC}
  L_{\rm \nu, IC} = \int^{\gamma_{\rm max}}_{\gamma_{\rm min}}
   P_{\rm IC}(\nu, \gamma_e) N(\gamma_e) d \gamma_e ,
\end{eqnarray}
  where 
\begin{eqnarray}
  \nonumber
   P_{\rm IC}(\nu, \gamma_e) = h \nu c 
    \int^{\infty}_0 n_{\rm ph}(\nu_{\rm s})
                     \sigma_{\rm IC}(\nu, \nu_{\rm s}, \gamma_e)
                     d \nu_{\rm s}.
\end{eqnarray}
   is the power spectrum for a single electron in isotropic photon fields. 
   Here,  $n_{\rm ph}(\nu_{\rm s})$ and
    $\sigma_{\rm IC}(\nu, \nu_{\rm s}, \gamma_e)$ are
   the number density of seed photons per unit frequency
   and the differential cross section for the IC scattering  given by 
  \citet{BG70}, which is valid both in the Thompson and KN regimes.
  The origins of the seed photons have been discussed in \S\ref{ele}.
  The number densities per unit frequency of these photons are then given as follows:
  $n_{\rm ph}(\nu_{\rm s}) = (U_{\rm ph} /  h \nu_{\rm ph}) \delta(\nu_s - \nu_{\rm ph})$ for
  the UV emissions from the accretion disc  
  ($U_{\rm ph} = U_{\rm UV},   \nu_{\rm ph} = \nu_{\rm UV}$),
  the IR emissions from the torus 
  ($U_{\rm ph} = U_{\rm IR}, \nu_{\rm ph} = \nu_{\rm IR}$),
  the NIR emissions from the host galaxy 
 ($U_{\rm ph} = U_{\rm V}, \nu_{\rm ph} =\nu_{\rm V}$),
  and CMB ($U_{\rm ph} = U_{\rm CMB}, \nu_{\rm ph} = \nu_{\rm CMB}$); 
  $n_{\rm ph}(\nu_{\rm s}) = L_{\rm \nu, lobe}/ h \nu_s 4\pi R^2 c$ for the lobe emissions.

%

%
%

\section{BROADBAND SPECTRUM OF THE SHELL}
\label{results}

 In the following we focus on powerful radio sources
 ($L_{\rm j} \sim 10^{45}-10^{47}{\rm ergs~s^{-1}}$)
 hosting  luminous quasars and present the temporal 
 evolutions of the energy distribution of non-thermal electrons
 and the resultant radiation spectra.
%
 
 As a fiducial case, we set the parameters for the ambient matter as 
$\rho_0 = 0.1 m_p~{\rm cm^{-3}}, r_0 =  1~{\rm kpc}$ and $\alpha = 1.5$,
 i.e., $\rho_a(r) = 0.1 m_p \rho_{0.1} (r / {\rm 1kpc})^{-1.5}{\rm cm^{-3}}$,
 where  $m_{p}$ is the proton mass and $\rho_{0.1} = \rho_0 / 0.1 m_p$. 
 These are indeed typical values for elliptical galaxies \citep[e.g.,][]{MZ98, MB03, FMO04, FBP06}. 
 We adopt ${\hat \gamma}_{\rm c} = 4/3$ and ${\hat \gamma}_{\rm a} = 5/3$.
 Then, $R$ is given from Eq. (\ref{R}) as 
\begin{eqnarray}
R(t) \approx        
  22              
                 {\rho}_{0.1}^{-2/7}
                  L_{45}^{2/7}
                   t_{7}^{6/7}~
                   {\rm kpc} ,      
\end{eqnarray}     
  where $t_{7} = t / 10^7~{\rm yr}$.
 As mentioned in  \S\ref{ele}, the power-law index, $p$,  
 for the injected non-thermal electrons (see Eq.~(\ref{Qinject})) is 
 fixed to $2$.
 Although the parameters $\xi$ and $\epsilon_e$ which
 characterize the electron acceleration efficiencies are highly uncertain,
 it is natural to expect that the range of these values are similar to those
 observed in the SNRs,
 since the nature of shocks considered here
 resembles those in the SNRs in that they are
 strong non-relativistic collision-less shocks driven into ISM.
 Here we take $\xi = 1$, which corresponds to the Bohm diffusion limit, 
 and $\epsilon_e = 0.01$ as illustrative values, based on the
 the observations of SNRs 
 \citep[e.g.,][]{DRB01, ESG01, BYU03, YYT04, SAH06, T08}.
 The luminosities
 of the emissions from accretion disc and dust torus
 are taken as
 $L_{\rm UV} = L_{\rm IR} = 10^{46}~{\rm ergs~s^{-1}}$
 for $L_{\rm j} > 10^{45}~{\rm ergs~s^{-1}}$,
 and $L_{\rm UV} = L_{\rm IR} = 10^{45}~{\rm ergs~s^{-1}}$ for
 $L_{\rm j} \leq 10^{45}~{\rm ergs~s^{-1}}$, while
 fixed value of $L_{\rm NIR} = 10^{45}{\rm ergs~s^{-1}}$
 is adopted for the emissions from the host galaxy.
 The parameters for the magnetic field and lobe emissions
 are chosen as
 $B= 10{\rm \mu G}$ 
 and
 $\eta = 0.01$. 
 The employed values of the parameters are summarized in Table \ref{tab1}.

\begin{deluxetable*}{lcccccccccccrrrrrrr}

\tabletypesize{\large}
\tablecaption{List of the Model Parameters.}

\startdata
\tableline\tableline

 Parameters &
 Symbols &
 Employed Values
 \\

\tableline

power-law index for density profile of ambient matter, $\rho_{a}(r)$&
 $\alpha$ &
 $1.5$
 \\
 
 reference position in $\rho_{a}(r)$&
 $r_0$ &
 $1~{\rm kpc}$
 \\

 mass density, $\rho_{a}(r_0)$, at $r=r_0$ &
 $\rho_0$ &
 $0.1 m_p~{\rm cm^{-3}}$ 
 \\

 magnetic field strength &
 $B$ &
 $10~{\rm \mu G}$
 \\

 luminosity of UV emissions from accretion disc &
 $L_{\rm UV}$ &
 $10^{46}~{\rm ergs~s^{-1}}$ $({\rm for}~ L_{\rm j}>10^{45}~{\rm ergs~s^{-1}})$
 \\

  &
  &
 $10^{45}~{\rm ergs~s^{-1}}$ $({\rm for}~ L_{\rm j}\leq 10^{45}~{\rm ergs~s^{-1}})$
 \\

 luminosity of IR emissions from dust torus &
 $L_{\rm IR}$ &
 $10^{46}~{\rm ergs~s^{-1}}$ $({\rm for}~ L_{\rm j}>10^{45}~{\rm ergs~s^{-1}})$
 \\

  &
  &
 $10^{45}~{\rm ergs~s^{-1}}$ $({\rm for}~ L_{\rm j}\leq 10^{45}~{\rm ergs~s^{-1}})$
 \\

 luminosity of NIR emissions from host galaxy&
 $L_{\rm NIR}$ &
 $10^{45}~{\rm ergs~s^{-1}}$
 \\

 ratio of lobe luminosity to jet power  &
 $\eta$ &
 $0.01$
 \\

 power-law index for energy distribution of injected electrons&
 $p$ &
 $2$
 \\

 gyro-factor &
 $\xi$ &
 1 \\

  energy fraction of non-thermal electrons &
 $\epsilon_e$ &
 0.01 

\enddata

\label{tab1}
\end{deluxetable*}

\subsection{Evolution of Non-thermal Electrons}
\label{ele2}

\begin{figure*}
 
\plotone{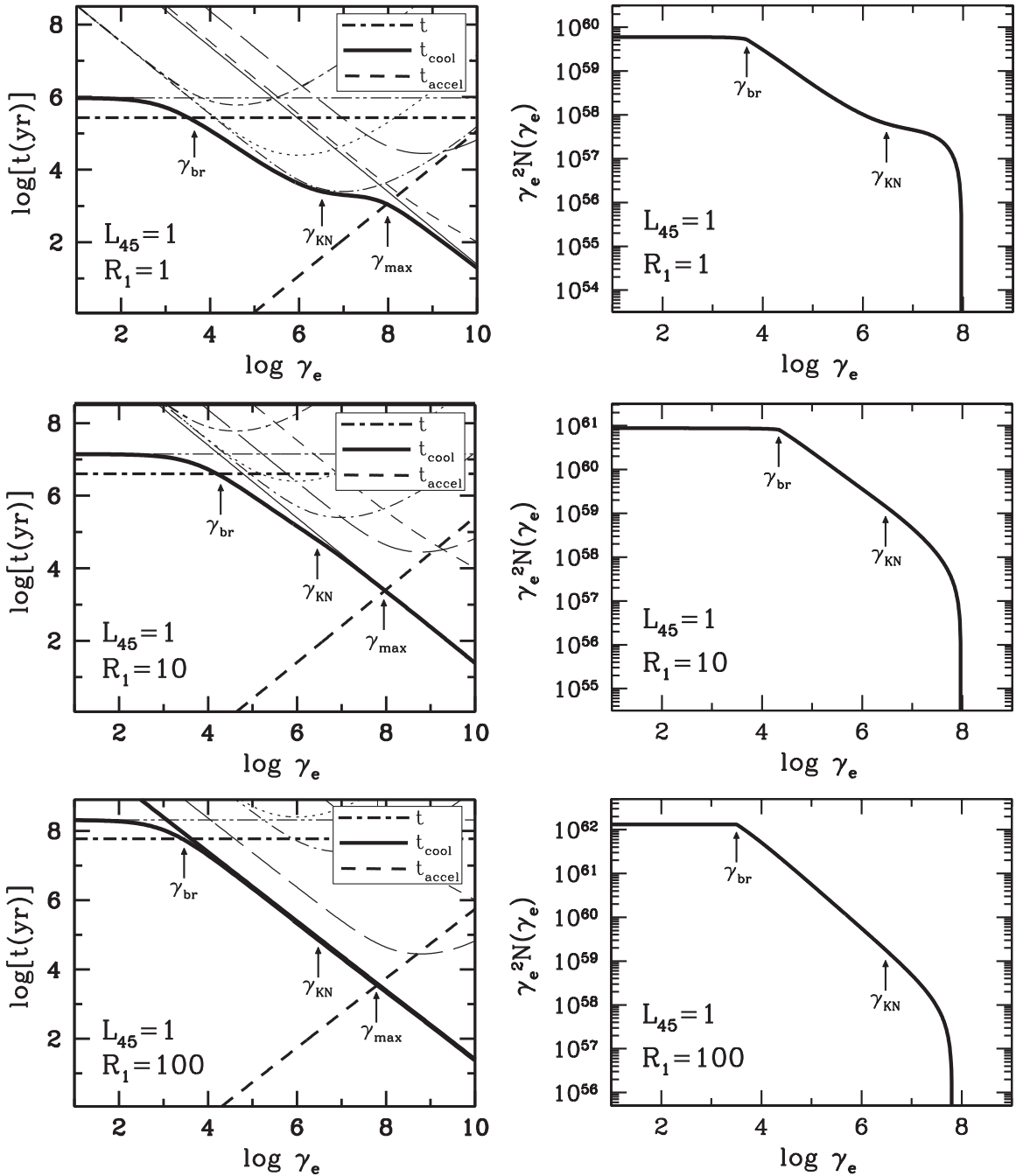}


\caption 
{Cooling and acceleration time scales 
 ({\it left panels}) and energy distributions 
 of non-thermal electrons ({\it right panels})
 for sources with the jet power of
 $L_{\rm j} = 10^{45}{\rm ergs~s^{-1}}$.
 The top, middle and bottom panels are shown for
 the source sizes of $R =1$, $10$ and $100~{\rm kpc}$,
 respectively.
 The thick lines in the left panels give the total cooling time 
 scale ({\it thick solid line}), the acceleration time scale 
 ({\it thick dashed line}) and the source age ({\it thick dot-dashed line}) 
 whereas the thin lines are contributions to the total 
 cooling time scale from various processes: adiabatic losses ({\it dot-dot-dashed line}),
 synchrotron emissions ({\it thin solid line}), and IC scatterings of UV disc photons 
 ({\it long-short-dashed line}), IR torus photons ({\it thin dot-dashed line}),
 NIR host galaxy photons ({\it dotted line}), lobe photons 
 ({\it thin short-dashed line}) and CMB ({\it thin long-dashed line}) photons.
  }
\label{tNL45}
\end{figure*}

\begin{figure*}
\plotone{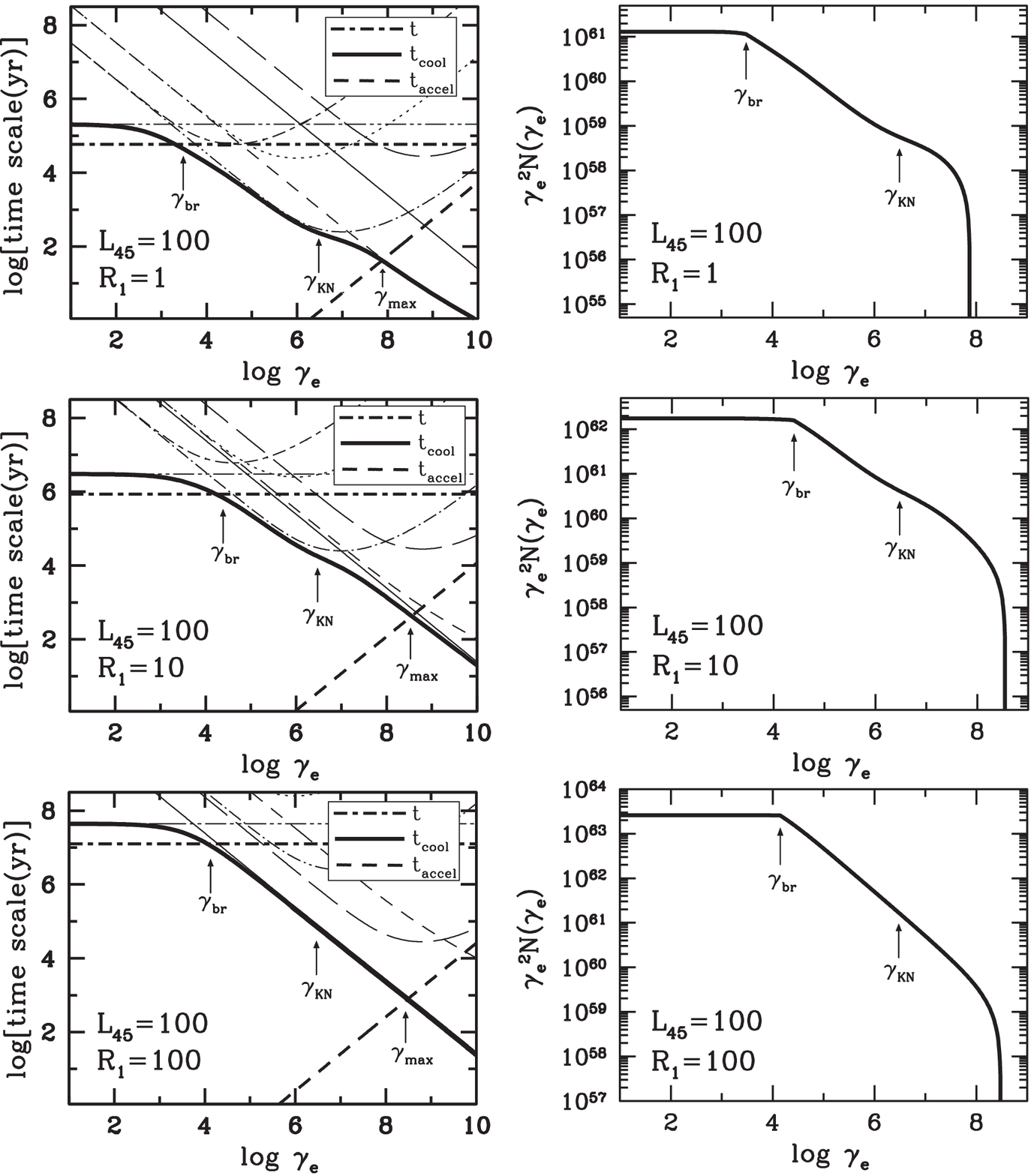}
\caption 
{Same as Fig. \ref{tNL45} but for the jet power of $L_{\rm j} = 10^{47}~{\rm ergs~s^{-1}}$.  } 
\label{tNL47}
\end{figure*}

 In Figs. \ref{tNL45} and \ref{tNL47},
 we display the cooling and acceleration time scales
 ({\it left panels}) and the  energy distribution of non-thermal electrons
 ({\it right panels}). The jet powers are chosen to be $L_{\rm j} = 10^{45}~{\rm ergs~s^{-1}}$ and  
 $L_{\rm j} = 10^{47}~{\rm ergs~s^{-1}}$ in Figs. \ref{tNL45} and \ref{tNL47}, respectively.
The top, middle and bottom panels in the figures are given for the source sizes of 
$R =1$, $10$, and $100~{\rm kpc}$,  respectively, which in turn corresponds to the 
different source ages.
%
%
%
 In addition to the total cooling time scale, $t_{\rm cool}$, 
 we show the contributions from the adiabatic loss,
 $t_{\rm ad} = \gamma_e / \dot{\gamma}_{\rm ad}$,
 synchrotron radiation, $t_{\rm syn} = \gamma_e / \dot{\gamma}_{\rm syn}$,
 and IC scatterings of the UV disc photons,
 $t_{\rm IC,UV} = \gamma_e / \dot{\gamma}_{\rm IC,UV}$,
 IR torus photons, $t_{\rm IC,IR} = \gamma_e / \dot{\gamma}_{\rm IC,IR}$,
 NIR host galaxy photons, $t_{\rm IC,NIR} = \gamma_e / \dot{\gamma}_{\rm IC,NIR}$,
 CMB photons, $t_{\rm IC,CMB} = \gamma_e / \dot{\gamma}_{\rm IC,CMB}$,
 and lobe photons, $t_{\rm IC,lobe} = \gamma_e / \dot{\gamma}_{\rm IC,lobe}$,
 together with the source age, $t$. 
%
%

 As mentioned in \S\ref{ele}, the Lorentz factor at the spectral break, $\gamma_{\rm br}$,
 corresponds to the electron  energy, above which cooling effects become important.
 It is hence determined roughly by the  condition $t \sim t_{\rm cool}$ as shown in the 
 left panels of Figs.~\ref{tNL45} and \ref{tNL47}.
 Since the adiabatic loss is dominant over the radiative coolings only below $\gamma_{\rm br}$,
 the latter is always more important when the coolings affect the
 electron distribution.
%
%
 The relative importance of the synchrotron emissions and IC emissions
 depends on the electron energy as well as the energy densities of 
 magnetic fields and photons.
 When the source is young and hence small, the energy
 loss is dominated by the IC emissions, since the energy density of 
 photons  is larger than that of magnetic fields (referred to as the IC-dominated stage).
 As the the source becomes larger, on the other hand, 
 the energy density of photons decreases ($U_{\rm ph} \propto R^{-2}$)
 and the synchrotron loss becomes more important (the synchrotron-dominated stage).  
%
  In the initial IC-dominated stage,
 the break Lorentz factor increases with the source size as 
 $\gamma_{\rm br} \propto R^{5/6}$, where we used the relations
 $t \propto R^{7/6}$ and
 $t_{\rm cool} \propto \gamma_e^{-1}
  U_{\rm ph}^{-1} \propto \gamma_e^{-1} R^2$.
 On the other hand,
  $\gamma_{\rm br}$ decreases with the source size
 as $\gamma_{\rm br} \propto R^{- 7/6}$ 
 in the synchrotron-dominated stage, since the relation 
 $t_{\rm cool} \propto \gamma_e^{-1}U_{\rm B}^{-1}
  \propto \gamma_e^{-1} R^{0}$ holds.
 Regarding the dependence on the jet power,
 for a given source size $R$, 
 $\gamma_{\rm br}$ increases with $L_{\rm j}$ roughly
 as $\gamma_{\rm br} \propto L_{\rm j}^{1/3}$
 because the dependencies of
 the source age
 and the radiative cooling time scale
 on 
 the jet power
 can be expressed
 as
 $t\propto L_{\rm j}^{- 1/3}$
 and 
 $\gamma_{\rm br} \propto \gamma_e^{-1}L_{\rm j}^{0}$,
 respectively. 


%
%
%
%

 Among the contributions to the IC losses, the scattering of the IR torus photons 
 is the largest at least in the IC-dominated stage thanks to the high energy density of the
 IR photons. Although the UV disc photons are assumed to have the same energy density as 
the IR photons ($U_{\rm UV} = U_{\rm IR}$), the cooling by the former is suppressed for 
 $\gamma_e \gtrsim m_e c^2 /4 h \nu_{\rm UV} \sim 1.3 \times 10^4$ by the KN effect 
 as seen in Figs. \ref{tNL45} and \ref{tNL47}. As a result, unless $L_{\rm UV}$ exceeds $L_{\rm IR}$
 significantly, most electrons with $\gamma_e \gtrsim 10^{4}$ cool
 predominantly through the IC scattering of IR photons.
 Then the transition 
 from the IC-dominated stage to the synchrotron-dominated stage
 occurs roughly at
\begin{eqnarray}
\label{stage}
 R_{\rm IC/syn} \sim 27 L_{\rm IR, 46}^{1/2} B_{-5}^{-2}{\rm kpc},
\end{eqnarray}
 which corresponds to the condition $U_{\rm IR} \sim U_{\rm B}$.
%
%
 Since the IC scattering of IR torus photons is also suppressed by the KN effect
 above the Lorentz factor given by 
\begin{eqnarray}
\label{KN}
\nonumber
 \gamma_{\rm KN} = m_e c^2 /4 h \nu_{\rm IR} \sim 3 \times 10^{6},
\end{eqnarray} 
 the coolings by the synchrotron emissions and/or the IC scattering of lobe photons 
 are important at the high energy end of non-thermal electrons  
 ($\gamma_{\rm KN} \lesssim \gamma_e \lesssim \gamma_{\rm max}$)
 for the source with $R < R_{\rm IC/syn}$ (see, e.g., the middle left panel of Fig. \ref{tNL45} and 
  the top and middle left panels of Fig. \ref{tNL47}).
For even larger sources with $R > R_{\rm IC/syn}$, the synchrotron emissions
are the dominant cooling mechanism at all energies.
 The contributions from the IC scatterings of CMB and host galaxy photons, on the other hand, 
 are modest at most during the whole evolution.
%

 The maximum Lorentz factor of the injected electrons,  
 which is determined by the condition $t_{\rm cool} = t_{\rm accel}$ 
 (or equivalently $\dot{\gamma}_{\rm cool} = \dot{\gamma}_{\rm accel}$),
 is $\gamma_{\rm max} \sim 10^7 - 10^9$ 
 for the source sizes $R \sim 1 - 100~{\rm kpc}$ and the jet
 power $L_{\rm j} \sim 10^{45} - 10^{47}~{\rm ergs~s^{-1}}$.
 Regarding the dependence on the source size,
 $\gamma_{\rm max}$ increases with $R$ 
 for compact sources in which 
 the cooling   at  $\gamma_e \sim \gamma_{\rm max}$ is dominated by
 the IC emissions.
 This is
 because the cooling time scale increases rapidly with size
 ($t_{\rm cool} \propto \gamma_e^{-1}F_{\rm KN}(\gamma_e)^{-1}U_{\rm ph}^{-1}
   \propto \gamma_e^{-1}F_{\rm KN}(\gamma_e)^{-1} R^{2}$),
 in contrast with the slow increase in acceleration time scale 
 ($t_{\rm accel} \propto \gamma_e \dot{R}^{-2} \propto \gamma_e R^{1/3}$).
 On the other hand,  
 for larger sources in which
 the cooling   at  $\gamma_e \sim \gamma_{\rm max}$ is dominated by
 the synchrotron emissions,
$\gamma_{\rm max}$ decreases slowly 
 with the source size
 ($\gamma_{\rm max} \propto R^{-1/6}$),
 since the
 cooling time scale is independent of the size
 ($t_{\rm cool} \propto \gamma_e^{-1} U_{\rm B}^{-1}
 \propto \gamma_e^{-1} R^{0}$).
%
 Note that
 this transition takes place earlier than the
 transition from the
 IC-dominated stage to the  synchrotron-dominated stage
 mentioned above
 owing to  the suppression of IC emissions
 in the  high energy electrons
 ($\gamma_e \gtrsim \gamma_{\rm KN}$)
 by the KN effect.
 For a given source size,
 higher values of $\gamma_{\rm max}$ are obtained
 for sources with larger jet powers, 
 since the acceleration time scale is shorter
 ($t_{\rm accel} \propto \gamma_e L_{\rm j}^{- 2/3}$).
%
%

 The resultant energy distributions of non-thermal electrons, $N(\gamma_e)$, are understood 
as follows. They have a sharp cut-off at $\gamma_e = \gamma_{\rm max}$ whereas 
 the original spectrum $N(\gamma_e) \propto \gamma_e^{-2}$ of the injected electrons 
is retained at low energies below $\gamma_{\rm br}$.
 Since the cooling effects are negligible, the energy spectrum for
 $\gamma_e \lesssim \gamma_{\rm br}$ are roughly given by
 $N(\gamma_e) \sim Q(\gamma_e)t$.
 The spectrum steepens for $\gamma_e \gtrsim \gamma_{\rm br}$ owing to 
the radiative cooling and is 
 roughly
determined by the energy dependence of $t_{\rm cool}$ 
as
 $N(\gamma_e) \sim Q(\gamma_e)t_{\rm cool}(\gamma_e)$.
 As mentioned above, for sources smaller than $R_{\rm IC/syn}$,
 the IC scattering of IR torus photons determines the spectral shape up to $\sim \gamma_{\rm KN}$,
 where the KN effect kicks in and hardens the energy spectra for higher energies
 (see, e.g., the top and middle panels of Figs. \ref{tNL45} and \ref{tNL47}).
 This feature is absent for larger sources ($R > R_{\rm IC/syn}$),
 because the synchrotron loss ($\dot{\gamma}_{\rm syn} \propto \gamma_e^2$)
 is dominant at all energies and the power is simply reduced by 1 ($N(\gamma_e) \propto \gamma_e^{-3}$) 
from the original spectrum of the injected electrons.
 For given source size,
 higher values of $N(\gamma_e)$ are obtained for sources with larger
 $L_{\rm j}$, 
 since the electron injection rate  $Q(\gamma_e)$ is higher.

\subsection{Evolution of Radiation Spectra}
\label{spectra2}

\begin{figure*}
\plotone{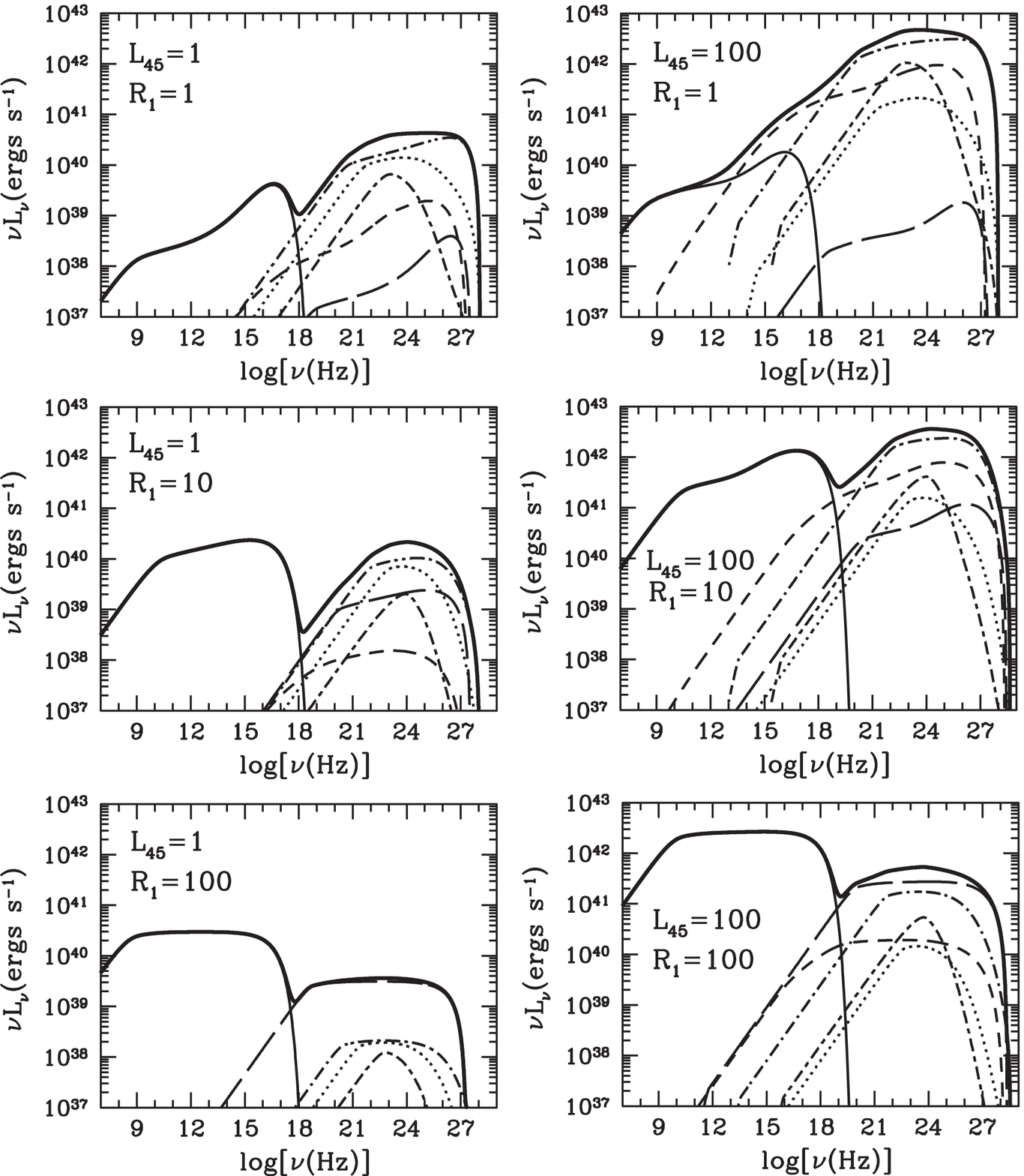}
\caption 
{
Broadband emission produced within the shell of
 sources with the jet powers of
 $L_{\rm j} = 10^{45}{\rm ergs~s^{-1}}$ ({\it left panels})
 and $L_{\rm j} = 10^{47}{\rm ergs~s^{-1}}$ ({\it right panels}).
 The top, middle and bottom panels are displayed for the source 
 sizes of $R =1$, $10$, and $100~{\rm kpc}$, respectively.
 The various lines show the contributions from the synchrotron 
 emissions ({\it thin solid line}) and IC scatterings of
 UV disc photons ({\it long-short-dashed line}),
 IR torus photons ({\it dot-dashed line}),
 NIR host galaxy photons ({\it dotted line}),
 CMB photons ({\it long-dashed line}) 
 and lobe photons ({\it short-dashed line}).
 The thick solid line is the sum of these emissions.
%
%
%
}
\label{Flux}
\end{figure*}

  In Fig. \ref{Flux}  we show the radiation spectra,
  $\nu L_{\nu}$,
  which are obtained by inserting into Eqs. (\ref{Lsyn}) and (\ref{LIC}) 
 the energy distributions of non-thermal electrons in the previous
  section.
 It is assumed that the jet powers
  are $L_{\rm j} = 10^{45}~{\rm ergs~s^{-1}}$ ({\it left panels}) and
   $L_{\rm j} = 10^{47}~{\rm ergs~s^{-1}}$ ({\it right panels}).
 The top, middle and bottom panels of the figure correspond to
 the source sizes of
 $R = 1~{\rm kpc}$, $10~{\rm kpc}$ and $100~{\rm kpc}$, respectively.
 In addition to the total luminosity ({\it thick solid line}),
 we show
 the contributions from the synchrotron emissions ({\it thin solid line})
 and the IC scatterings of UV disc photons ({\it long-short-dashed line}),
 IR torus photons ({\it dot-dashed line}), NIR host galaxy photons ({\it dotted line}),
 CMB photons ({\it  long-dashed line}) and lobe photons ({\it short-dashed line}).

 The synchrotron emissions are the main low-frequency component,  
 which extends from radio to X-ray $\sim 2 (\gamma_{\rm max}/10^8)^2 B_{-5}$ ${\rm keV}$.
 The IC emissions become remarkable at higher frequencies 
 up to gamma-ray $\sim 50(\gamma_{\rm max}/10^8)~{\rm TeV}$.
 As mentioned in \S\ref{ele2},
 the dominant radiative process changes from the
 IC emissions to the synchrotron radiations
 as the source becomes larger.
 As a result, the former is more luminous than the latter  
 when the source is young and compact
 ($R \lesssim R_{\rm IC/syn}$) and vice versa.
 Owning to the hardening of the energy distributions of non-thermal electrons
 at $\gamma_e \gtrsim \gamma_{\rm KN}$
 (see Figs. \ref{tNL45} and \ref{tNL47})
 for compact sources ($R\lesssim R_{\rm IC/syn}$), their synchrotron spectrum 
 becomes also harder than for larger sources.
 This feature is not so remarkable for most of the IC components 
 mainly because the emissions themselves are suppressed by the KN effect.
%
%
 The IC scattering of IR dust-torus photons 
 dominates over other IC components
 up to the source size of $R \sim 85 L_{\rm IR, 46}^{1/2}{\rm kpc}$.
 For larger sources, on the other hand,  
 the IC scattering of CMB photons becomes more important, since CMB has the largest 
 energy density of all photons considered in this paper.
%
 The contributions from the UV disc photons and NIR host galaxy photons
 are modest at best through the entire evolution.

 The energy injection rate above $\gamma_{\rm br}$
is roughly equal to the radiative output  ($\gamma_e^2 m_e c^2 Q(\gamma_e) \sim \nu L_{\nu}$) 
of non-thermal electrons in this energy regime 
 because the cooling time scale is shorter than the dynamical time scale. 
 Since the energy injection rate is independent of the electron energy 
 ($\gamma_e^2 Q(\gamma_e) \propto \gamma_e^0$),
  these non-thermal electrons produce a rather flat and broad spectrum
 ($\nu L_{\nu} \propto \nu^{0}$) in the corresponding frequency range.
 From the relation
 $\gamma_e^2 m_e c^2 Q(\gamma_e) =
  K m_e c^2 \sim  0.1 \epsilon_e L_{\rm j} / [{\rm ln}(\gamma_{\rm
 max})]$
 (see \S\ref{ele}),
 we can give a rough estimate to
 the peak luminosity as
\begin{eqnarray}
 (\nu L_{\rm \nu})_{\rm peak}
  \sim  4.0 \times 10^{40} \epsilon_{-2}L_{45} 
~{\rm ergs~s^{-1}},
\label{Lpeak}
\end{eqnarray}
 where $\epsilon_{-2} = \epsilon_e / 0.01$. 
 In the above equations,
 we ignored the dependence of luminosity 
 on $\gamma_{\rm max}$
 because the luminosity
  only scale logarithmically with its value,
 and employed a typical value  $\gamma_{\rm max} \sim 10^{8}$.
 The feature is clearly seen in Fig. \ref{Flux}. 
Indeed, the spectra are flat with
 a peak luminosity given approximately by Eq. (\ref{Lpeak}).
 As mentioned in \S\ref{ele},
 the emission luminosity 
 scale approximately linearly
 with the acceleration efficiency $\epsilon_e$ and  the
 jet power $L_{\rm j}$.
 For given values of $\epsilon_e$ and $L_{\rm j}$,
 while the value of $(\nu L_{\nu})_{\rm peak}$ remains nearly constant,
the frequency range, where the spectrum is flat, varies with the source size
because of the changes in $\gamma_{\rm br}$, $\gamma_{\rm max}$  
and the main emission mechanism (synchrotron or IC). 
 It is emphasized that the peak luminosity
 is chiefly governed by $\epsilon_e$ and $L_{\rm j}$
 and is quite insensitive to
 the  magnetic field strength and seed photons, which will
 only affect the frequency range of the flat spectrum.
%
 This means that if $L_{\rm j}$  is constrained by other independent methods \citep[e.g.,][]{AFT06, IKK08},
 the observation of the peak luminosity will enable us to obtain information on the highly unknown 
acceleration efficiency $\epsilon_e$.

%
%
%
%

\section{COMPARISON WITH EMISSIONS FROM THE LOBE} 
\label{comp}

 Despite our focus on the non-thermal emission from the 
 shell, emission from the lobe (or, cocoon) 
 is an another important ingredient which arises as 
 a consequence of interaction of jet with ambient medium.
 As shown by \citet{SBM08},
 lobes can produce very bright emissions in broadband.
 Therefore, in order to
 consider the application of our model to the observation
 of radio sources,
 it is essential to investigate the
 relative significance of the two emissions,
 since the lobe component may hamper the detection.
 In this section, we evaluate the emissions from the lobe
 and show the quantitative comparison between the spectra of
 the lobe and shell.
%



\subsection{Model for Emissions from the Lobe}

 We basically follow the same procedure  employed for the shell
 in calculating
 the energy distribution of the electrons 
 and the resulting emissions.
 Since we are considering 
 the electrons residing in the lobe,
 the energy stored in the cocoon,
 $E_{\rm c} = P_{\rm c}V_{\rm c}/(\gamma_{\rm c} - 1)$,
 is the energy budget for the emissions.
 From the dynamical model described in \S\ref{dynamics},
 the total internal energy stored in the cocoon is given by
\begin{eqnarray}
  E_{\rm c} = f_{\rm lobe}L_{\rm j} t,
\end{eqnarray}
 where 
 $f_{\rm lobe} = (5 - \alpha) (7 - 2 \alpha)/
 [2 \alpha^2 + (1 - 18\hat{\gamma}_{\rm c})\alpha + 63 \hat{\gamma}_{\rm c} - 28] $. 
 For ${\hat \gamma}_{\rm c} = 4/3$, 
  ${\hat \gamma}_{\rm a} = 5/3$ and
 $\alpha = 1.5$,
 $f_{\rm lobe} = 7/13 \sim 0.5$ is obtained.
 As can be seen from the above equation and  Eq. (\ref{Es}),
 $E_{\rm c}$ is larger than $E_{\rm s}$ by a factor of $\sim 5$.

 The energy distribution of the electrons
 in the lobe is determined based on Eq. (\ref{kinetic})
 as in the case of the shell.
 As described in \S\ref{ele},
 by evaluating the electron injection rate, $Q(\gamma_e)$,
 and the cooling rate,  $\dot{\gamma}_{\rm cool}(\gamma_e)$,
 the energy distribution of the electrons is obtained by
 putting these quantities in Eq. (\ref{N}) which corresponds to
 the solution of Eq. (\ref{kinetic}).
 In the present study,
 we assume that the electrons injected
 into the lobe
 have a power-law energy distribution 
 given as 
\begin{eqnarray}
\nonumber
  Q(\gamma_e) = K_{\rm lobe} \gamma_e^{- p_{\rm lobe}} 
 {\rm for}~~  \gamma_{\rm min, lobe} \leq \gamma_e
                       \leq \gamma_{\rm max, lobe}.
\end{eqnarray}
 While
 the  employed values of the power-law index 
 and minimum Lorentz factor are the same as those
 adopted for the shell
 ($p_{\rm lobe} = 2$ and $\gamma_{\rm min, lobe} = 1$),
 the
 maximum Lorentz factor is fixed as
 $\gamma_{\rm max, lobe} = 10^5$. 
  We will comment on the assumed value of $\gamma_{\rm max, lobe}$
  later in \S\ref{vs}.
%
 The normalisation factor, $K_{\rm lobe}$,
 is
 determined from the assumption 
 that a fraction, $\epsilon_{e, {\rm lobe}}$,
 of the energy deposited in the lobe is carried by
 non-thermal electrons: 
 $\int^{\gamma_{\rm max, lobe}}_{\gamma_{\rm min, lobe}} 
   (\gamma_e - 1) m_e c^2 Q(\gamma_e) d\gamma_e=
 \epsilon_e dE_{\rm c}/dt = f_{\rm lobe} \epsilon_{e, {\rm lobe}} L_{\rm j}$.
 A rough estimation of the normalization factor is 
 obtained as
 $K_{\rm lobe} \sim 0.5 \epsilon_{e, {\rm lobe}} L_{\rm j} /
 [m_e c^2{\rm ln}({\gamma_{\rm max, lobe}})]$,
 where we used $f_{\rm lobe} \sim 0.5$.
 Regarding the cooling rate of the electrons, 
 we take into account the adiabatic losses
 as well as radiative losses due to 
 the synchrotron and IC emissions.
 The adiabatic cooling rate is evaluated from Eq. (\ref{adiabatic})
 and coincides with that of the shell.
%
 The synchrotron cooling rate
 is determined from  Eq. (\ref{synchro})
 under the assumption that a fraction  
 $\epsilon_{\rm B}$, of the energy $E_{\rm c}$ is 
 carried by the magnetic fields. 
 The corresponding energy density of magnetic field is given as
\begin{eqnarray}
 \nonumber
 U_{\rm B}  = 
 \frac{ \epsilon_{\rm B} E_{\rm c}}{  V_{\rm c}} 
  \approx  4.0 \times 10^{-9} \epsilon_{{\rm B}, 0.1} 
 \rho_{0.1}^{1/3} L_{45}^{2/3} R_{1}^{-11/6}
 {\rm erg~cm^{-3}},
\end{eqnarray}
 where  $\epsilon_{{\rm B}, 0.1} = \epsilon_{\rm B} / 0.1$.
%
%
 As for the IC cooling rate, 
 we take into account 
 all seed photon fields which are considered
 in the shell 
 except for the synchrotron photons from the lobe.
 Considering the IC of lobe photons,
 although a particular 
 spectrum
 was assumed
 in the case of the shell (see \S\ref{ele}),
 to be self-consistent,
 the distribution of the 
 seed photons should be determined
 from the calculated synchrotron emissions
 (synchrotron self-Compton).  
 However,
 the method given in \S\ref{ele}
 used to solve Eq. (\ref{kinetic})
 cannot be applied when a seed photon field
 has dependence on the electron distribution.
 Therefore,
 since the cooling rate due the 
 synchrotron self-Compton is modest at best in any case considered
 in the present study,
 we  neglected its effect 
 merely for simplicity.
%
 As in the case of the shell,
 we assume that 
 the photons
 from the disc, torus, host galaxy and CMB   
 are monochromatic with frequencies given in \S\ref{ele}.
 On the other hand, 
 while same value is employed for CMB, 
 the energy densities of
 the  disc,  torus and  host galaxy photons
 are taken to be  larger than those 
 in the shell by a factor of $3$,
 since the lobe is located closer to the emitting sources.
 The cooling rate due the IC of these photons
 are evaluated from Eq. (\ref{IC})
 based on the above mentioned  spectrum and energy densities.

 The emission spectrum of the lobe is calculated
 by following the procedure described in
 \S\ref{spectra}.
 From  the obtained energy distribution of the non-thermal electrons,
 spectra of the synchrotron and IC emission are 
 evaluated from Eqs. (\ref{Lsyn}) and (\ref{LIC}), respectively.
%
 The magnetic field and seed photon fields considered above is
 used for the calculation. 
 In addition,
 although we did not consider its contribution
 on the cooling rate,
 we also evaluate the IC of the  photons from the lobe
 based on the calculated synchrotron spectrum.

\subsection{Lobe vs Shell}
\label{vs}

\begin{figure*}
\plotone{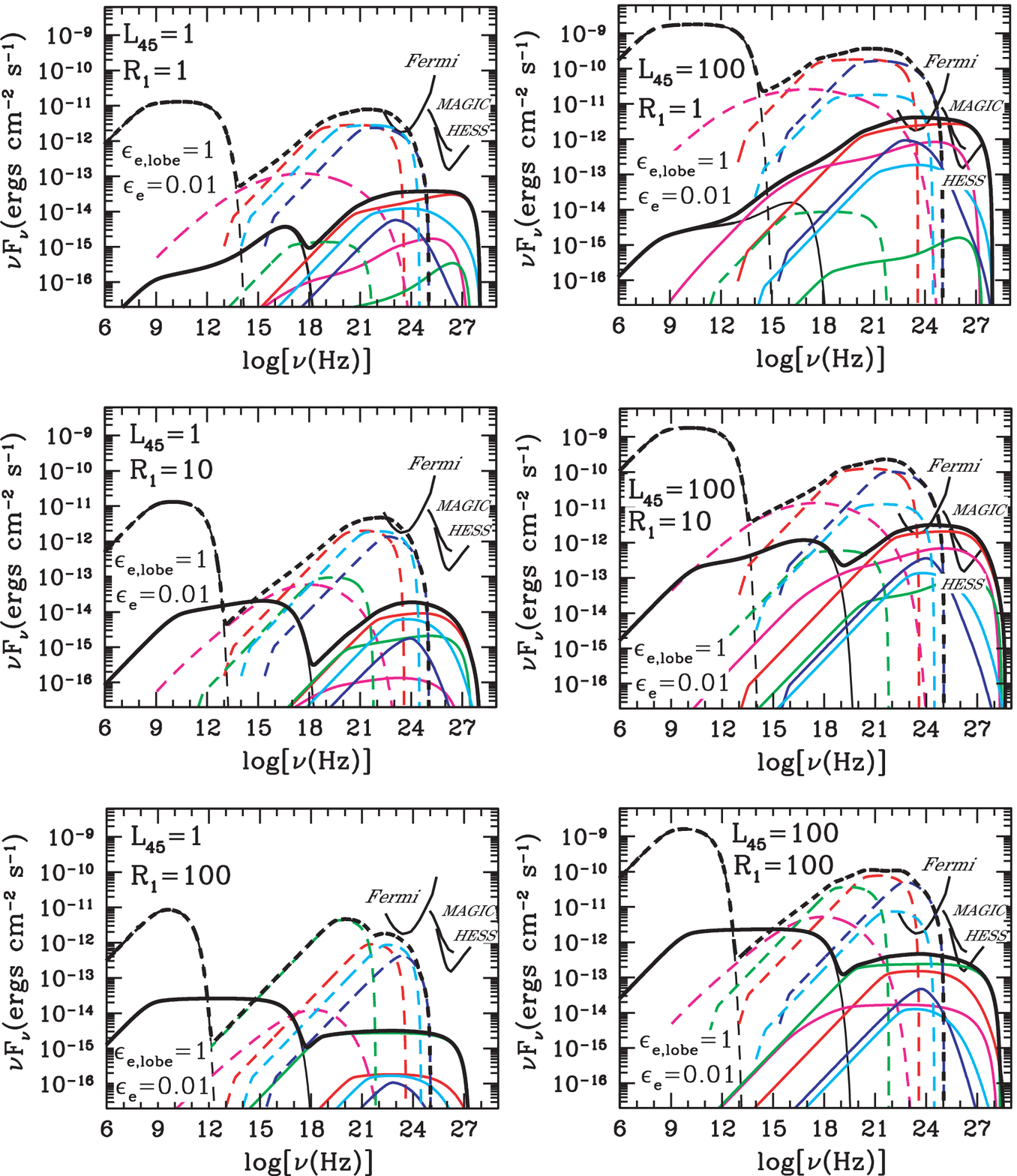}
\caption 
{
Photon fluxes from sources with the jet power of
 $L_{\rm j} = 10^{45}{\rm ergs~s^{-1}}$ ({\it left panels}) and
 $L_{\rm j} = 10^{47}{\rm ergs~s^{-1}}$ ({\it right panels})
 located at the distance of $D=100~{\rm Mpc}$.
 The top, middle and bottom panels are displayed for the source 
 sizes of $R =1$, $10$, and $100~{\rm kpc}$, respectively.
 The dashed and solid lines display the photon fluxes 
 from  the lobe and shell, respectively.
 The various lines show 
 the contributions from the synchrotron emissions ({\it thin black line})
 and the IC scatterings of UV disc photons ({\it blue line}),
 IR torus photons ({\it red line}),
 NIR host galaxy photons ({\it light blue line}),
 CMB photons ({\it  green line}) and
 lobe synchrotron photons ({\it purple line}). 
 The thick black lines are the the sum of these fluxes.
%
 Also shown are the sensitivities of the Fermi,
  MAGIC and HESS telescopes.
 }
\label{Fluxcs}
\end{figure*}

 For an illustrative purpose,
 here we examine 
 the emissions from the lobe by assuming
 $\epsilon_{e, {\rm lobe}} = 1$,
 an extreme case where
 all energy stored in the cocoon is converted to that of the
 non-thermal electrons, and
 $\epsilon_{\rm B} = 0.1$, a magnetic field strength factor of 
 few below the equipartition value.
 These values are identical to those employed in  \citet{SBM08}.
  In Fig. \ref{Fluxcs}  we show the obtained photon fluxes,
  $\nu F_{\nu}$,
  from the lobe. 
 In order to illustrate the comparison
 between the emissions from the lobe and the shell,
 photon fluxes from the shell are also shown in the figure.
 Same set of parameters
 which was previously assumed (Table. \ref{tab1})
 is employed for the emissions from the shell.  
 It is assumed that the jet powers
 are $L_{\rm j} = 10^{45}~{\rm ergs~s^{-1}}$ ({\it left panels}) and
 $L_{\rm j} = 10^{47}~{\rm ergs~s^{-1}}$ ({\it right panels}) and
 the source is    located at a distance of $D = 100~{\rm Mpc}$.
 The top, middle and bottom panels of the figure correspond to the source sizes of
 $R = 1~{\rm kpc}$, $10~{\rm kpc}$ and $100~{\rm kpc}$, respectively.
 We also plot the sensitivities of the Fermi satellite
 (http://www-glast.stanford.edu/)
 and HESS
 (http://www.mpi-hd.mpg.de/hfm/HESS/)
 and MAGIC
(http://magic.mppmu.mpg.de/)
 telescopes.
%
%
 The dashed and solid lines display the photon fluxes 
 from  the lobe and shell, respectively.
 In addition to the total photon flux ({\it thick black line}),  
 the contributions from the synchrotron emissions ({\it thin black line})
 and the IC scatterings of UV disc photons ({\it blue line}),
 IR torus photons ({\it red line}),
 NIR host galaxy photons ({\it light blue line}),
 CMB photons ({\it  green line}) and
 lobe synchrotron photons ({\it purple line})
 are also shown.

%
 As is seen in  the figure,
 IC emissions from the lobe
 tend to be brighter
 when the source size is smaller 
 as in the case of the shell.
 On the other hand, however,
 the  emissions are  synchrotron-dominated
 even for compact sources ($R\sim 1-10~{\rm kpc}$)
 since the magnetic field strength
 is larger than that of the shell.
 As in the case of the shell,
 the  luminosity
 scales linearly with $\epsilon_{e, {\rm lobe}}$
 and $L_{\rm j}$  (see \S\ref{spectra}). 
 A rough estimate to the peak luminosity is obtained as
\begin{eqnarray}
 (\nu L_{\rm \nu})_{\rm peak, lobe}
  \sim  2.0 \times 10^{43} \epsilon_{e, {\rm lobe}}L_{45} 
~{\rm ergs~s^{-1}}.
\label{Lpeakl}
\end{eqnarray}
 It should be noted that
 the 
 photon fluxes 
 displayed in the figure correspond to
 an upper limit since $\epsilon_{e, {\rm lobe}} = 1$ is assumed.

 From Eqs. (\ref{Lpeak}) and (\ref{Lpeakl}), 
 the ratio between the
 peak luminosities of 
 the emissions from the lobe and shell
 is obtained as
 $ \sim 5 \epsilon_{e, {\rm lobe}} / \epsilon_e$.
 Hence, the contrast between the two emissions is
 determined by the parameters
 $\epsilon_{e, {\rm lobe}}$ and $\epsilon_e$.
%
 If $\epsilon_{e, {\rm lobe}}$ is
 much larger than $\epsilon_e$,
 the overall spectra is dominated by the lobe.
 Although there is little constraint on the values of
 $\epsilon_{e, {\rm lobe}}$ and $\epsilon_e$,
 it is  expected that the
 ratio $\epsilon_{e, {\rm lobe}} / \epsilon_e$ is
 indeed large in most of the radio sources as is examined here
 due to the fact that, while large number of radio observations identified
 non-thermal emissions  from the lobe,
 no clear evidence of the shell emissions have been reported 
 so far at radio.
%
%
 It is emphasized, 
 however, that, even in the case where $\epsilon_{e, {\rm lobe}}$ is
 significantly larger than $\epsilon_e$,
 there are frequency ranges in which the emissions from the shell
 can overwhelm those from the lobe as is seen in Fig. \ref{Fluxcs}.
 This is
 due to the difference
 between the maximum energy of the electrons
 in the lobe and shell.
 Since
 $\gamma_{\rm max}\sim 10^8$ largely exceed $\gamma_{\rm max, lobe}= 10^5$,
 the synchrotron and IC emissions from the shell
 extend up to much higher frequencies
 than those from the lobe, and, as a result, 
 the shell can dominate the emission spectra
 at the frequencies
 above the 
 cut-off frequencies of the synchrotron  and IC emissions
 from the lobe.

 Regarding the synchrotron emission, 
 although the high frequency end of the emission is overwhelmed by 
 the  low frequency tail of the IC emissions from the lobe,
 emissions from the shell can become dominant 
 at frequencies  above the cut-off frequency of synchrotron emission
 from the lobe which is roughly given as
 $h \nu \sim 6\times 10^{-2} (\gamma_{\rm max, lobe} / 10^5)^2
 \epsilon_{\rm B, 0.1}^{1/2} \rho_{0.1}^{1/6}
 L_{45}^{1/3}R^{-11/12}~ {\rm eV}$
 when the source size is large ($R \gtrsim R_{\rm IC/syn}$).
 For example,
 emissions from the shell can be observed at frequencies from IR to optical
 without strong contamination from the lobe
 for sources with size of $R \sim 100~{\rm kpc}$
 (see the bottom panels of Fig. \ref{Fluxcs}).
 Hence, deep observations of radio galaxies
 at these frequencies may lead to discovery of the shell emissions.
 It is noted that a broader range of frequencies could be 
 detectable if the ratio $\epsilon_{e, {\rm lobe}} / \epsilon_{e}$
 is smaller.
 For compact sources ($R \lesssim R_{\rm IC/syn}$), however,
 synchrotron emissions are likely to be completely overwhelmed by the 
 lobe components, 
 since the synchrotron emissions from the shell is strongly suppressed
 (see the top panels of Fig. \ref{Fluxcs}).

 On the other hand,
 IC emissions from the shell are not subject to 
 the contamination from the lobe 
 irrespective to the size of the source
 at frequencies
 above the cut-off frequency of IC emissions 
 from the lobe which is roughly given as $h \nu \sim
 \gamma_{\rm max, lobe} m_e c^2
 \sim 50 (\gamma_{\rm max, lobe}/10^5)~{\rm GeV}$,
 since the lobe emissions cannot extend above the frequency.
 It is also noted that
 compact sources ($R \lesssim R_{\rm IC/syn}$) are favored for detection, 
 since the luminosity is higher. 
 Contrary to the above argument,
 although the same value was assumed for
 the maximum Lorentz factor
 ($\gamma_{\rm max, lobe} = 10^5$), 
 \citet{SBM08} argued
 that the lobe can produce prominent emissions up to $\sim {\rm TeV}$
 gamma-ray.
 It should be noted, however, that the emissions
 above the frequency  $h \nu \sim 10 {\rm GeV}$ in their study 
 is  considerably overestimated since the KN effect was not
 taken into account. 
 Hence, 
 due to the absence of lobe emissions, 
 our study demonstrates that the 
 energy range of $h \nu \gtrsim 10~{\rm GeV}$ 
 is the most promising window for the detection of the shell emissions.
 This can be confirmed in  Fig. \ref{Fluxcs} indeed.
 For example,
 while the detection of $\sim {\rm GeV}$ gamma-rays
 by the Fermi telescope may be hampered by the 
  emissions from the lobe,
 $\sim {\rm TeV}$ gamma-rays are
  accessible to the 
 MAGIC, HESS and, although not displayed in the figure, also
 VERITAS gamma-ray telescopes 
 for the most powerful source with the jet power of
 $L_{\rm j} = 10^{47}~{\rm ergs~s^{-1}}$
 located at  $D = 100~{\rm Mpc}$.
 It is worth noting that the maximum Lorentz factor $\gamma_{\rm max}$
 of non-thermal electrons can  be constrained by the detection of  
 high energy cut-off at $h \nu \sim 50(\gamma_{\rm max}/10^8)~{\rm TeV}$ 
 in the IC emissions. 
 Hence, the observation of the high energy gamma-rays can provide us
 with information not only on $\epsilon_e$ as mentioned
 in \S\ref{spectra2}
 but also on the gyro-factor $\xi$.

 Lastly, let us comment on the employed values of
 $\gamma_{\rm max}$ and $\gamma_{\rm max, lobe}$.
 In the above discussions, we have shown that, 
 even in the case of $\epsilon_{e, {\rm lobe}} \gg \epsilon_e$,
 the high frequency part of
 synchrotron ($\sim$ IR $-$ optical) and IC emissions
 ($\sim 10~{\rm GeV} - 10~{\rm TeV}$) 
 from
 the shell can overwhelm the lobe emissions.
 The above argument will be valid as long as
 (i) $\gamma_{\rm max} \gtrsim 10^7$ and
 (ii) $\gamma_{\rm max, lobe} \lesssim 10^5$ are satisfied.
 The condition (i) holds if the acceleration 
 at the shell takes place in  nearly Bohm limit ($\xi \sim 1$)
 as shown in Figs. \ref{tNL45} and \ref{tNL47}.
 As mentioned in \S\ref{results},
 since the property of the shock considered 
 here is similar to those of the SNRs
 in  which electron acceleration at nearly Bohm diffusion
 is inferred 
 \citep[e.g.,][]{YYT04, SAH06, T08},
  it is natural to expect that this is also the case
 for the shell.
 In fact,
 $\xi \sim 1$ inferred in the shell associated with radio galaxy
 Centaurus A  from the X-ray observations \citep{CKH09}.
 Hence, we expect that condition (i) is satisfied
 for the sources  we focus here. 
%
 Regarding the value of $\gamma_{\rm max, lobe}$,
 multi-wavelength studies of                     
 powerful FRII radio galaxies  show that maximum Lorentz factor
 of electrons residing in the lobe
 does not extend well beyond
 $\sim 10^5$ \citep{SCH07, G09, YTI10}
\footnote{ For ``low-power'' FRII radio sources
 with hotspots  showing
 synchrotron emissions above optical frequencies, however, 
 electrons may be accelerated to higher energies \citep{BMP03, HHW04}.}.
 Although the detail of the acceleration process is highly uncertain,
 if this feature is also common in  compact sources ($R \sim 1-10~{\rm kpc}$),
 condition (ii) is likely to be satisfied
 irrespective to the source size.
 It is worth to note that the
 recent study by \citet{OMS10}
 has shown that
 the broadband spectra in some compact radio sources
 can indeed
 be  fitted under the assumption of $\gamma_{\rm max, lobe} \sim 10^5$.

\section{SUMMARY}
\label{summ}

 We have explored the temporal evolution of the emissions by accelerated
 electrons in 
 the shocked shell produced by AGN jets. Focusing on the powerful
 sources which
 host
 luminous quasars, we have calculated the spectra of the synchrotron emission as well as 
 the IC scatterings of various photons that will be relevant in this context.  
 We have used a simple analytic model that describes
 the dynamics of the expanding  shell
 and estimated the energy distribution of non-thermal electrons based on 
this model, taking properly into account both the adiabatic and radiative coolings.
 Below we summarize our main findings in this study.

 \vspace{2mm}
 
 1.
 When the source is small ($R \lesssim R_{\rm IC/syn} \sim 27 L_{\rm IR, 46}^{1/2} B_{-5}^{-2}{\rm kpc}$),
 the dominant radiative process is the IC scattering of IR photons emitted 
 from the dust torus.
 For larger sources, on the other hand, the synchrotron emissions dominate 
over the IC emissions, since the energy density of photons becomes smaller than 
that of magnetic fields ($U_{B} > U_{\rm ph} \propto R^{-2}$).
Through the entire evolution, the spectrum is rather broad and flat, and 
the peak luminosity is approximately given by 
 $(\nu L_{\nu})_{\rm peak} \sim  4.0 \times 10^{40} \epsilon_{-2}L_{45}~{\rm ergs~s^{-1}}$, 
since it is roughly equal to the energy injection rate, which is in turn determined 
by the jet power $L_{\rm j}$ and acceleration efficiency $\epsilon_e$.

 \vspace{2mm}

 2.
 The broadband spectra extend from radio up to 
  $\sim 10~{\rm TeV}$ gamma-ray energies
 for a wide range of source size ($R\sim 1-100~{\rm kpc}$)
 and jet power ($L_{\rm j}\sim 10^{45}-10^{47}~{\rm ergs~s^{-1}}$).
 By comparing the emissions with those from the lobe,
 we find that 
 the synchrotron emissions at IR and optical frequencies
 can be observed without being hampered by the lobe emissions
 for extended sources ($R \gtrsim R_{\rm IC/syn}$),
 while the IC emissions at $h \nu \gtrsim 10~{\rm GeV}$ 
 can be observed with the absence of contamination from the lobe 
 irrespective of the source size.
 In particular, it is predicted that,
 for most powerful nearby sources  ($L_{\rm j} \sim 10^{47}~{\rm ergs~s^{-1}}$, 
$D \lesssim 100~{\rm Mpc}$), $\sim {\rm TeV}$  gamma-rays produced via
the IC emissions can be detected by the modern Cherenkov telescopes
 such as MAGIC, HESS and VERITAS. 
%


\acknowledgments

We are grateful to H. Nagai and M. Tanaka for useful comments and
discussions.
We thank the anonymous referee for constructive comments.
This study was supported by Program for Improvement of Research
Environment for Young Researchers from Special Coordination Funds for
Promoting Science and Technology (SCF) commissioned by the Ministry of
Education, Culture, Sports, Science and Technology (MEXT) of Japan.
 This work was partially supported by the Grants-in-Aid for the
 Scientific Research (17540267, 19104006, 21540281) from Ministry of Education,
 Science and Culture of Japan. 
This work is supported in part by Ministry of Education, 
Culture, Sports,Science, and Technology (MEXT) 
Research Activity Start-up 2284007 (NK).

\begin{thebibliography}{}


\bibitem[Allen et al.(2006)]{AFT06} Allen, S.~W., Dunn, 
R.~J.~H., Fabian, A.~C., Taylor, G.~B., 
\& Reynolds, C.~S.\ 2006, \mnras, 372, 21 


\bibitem[Bamba et al.(2003)]{BYU03} Bamba, A., Yamazaki, R., 
Ueno, M., \& Koyama, K.\ 2003, \apj, 589, 827 



\bibitem[Begelman et al.(1984)]{BBR84} Begelman, M.~C., 
Blandford, R.~D., \& Rees, M.~J.\ 1984, Reviews of Modern Physics, 56, 255 



\bibitem[Begelman \& Cioffi(1989)]{BC89} Begelman, M. C.,
    \& Cioffi, D. F.  1989,  \apj, 345, 21


\bibitem[Bell(1978)]{B78}
Bell, A. R. 1978,
\mnras, 182, 443


\bibitem[Berezhko(2008)]{B08} Berezhko, E.~G.\ 2008, \apjl, 
684, L69 





\bibitem[Blandford \& Eichler(1987)]{BE87} 
Blandford, R. D., \& Eichler, D. 1987, Phys. Rep., 154, 1


\bibitem[Blumenthal 
\& Gould(1970)]{BG70} Blumenthal, G.~R., \& Gould, R.~J.\ 1970, Reviews of Modern Physics, 42, 237 




\bibitem[Brunetti et al.(2003)]{BMP03} Brunetti, G., Mack, 
K.-H., Prieto, M.~A., \& Varano, S.\ 2003, \mnras, 345, L40 



\bibitem[Carilli et al.(1988)]{CPD88} Carilli, C.~L., Perley, 
R.~A., \& Dreher, J.~H.\ 1988, \apjl, 334, L73 




\bibitem[Carilli  \& Taylor(2002)]{CT02} 	
 Carilli, C. L., \& Taylor, G. B. 2002, ARA\&A, 40, 319


\bibitem[Castor et al.(1975)]{CMW75} Castor, J., McCray, R., 
\& Weaver, R.\ 1975, \apjl, 200, L107 



\bibitem[Clarke et al.(2001)]{CKB01} Clarke, T.~E., Kronberg, 
P.~P., \& B\"ohringer, H.\ 2001, \apjl, 547, L111 





\bibitem[Croston et al.(2009)]{CKH09} Croston, J.~H., et al.\ 
2009, \mnras, 395, 1999 



\bibitem[de Ruiter et 
al.(2005)]{RPC05} de Ruiter, H.~R., Parma, P., Capetti, A., Fanti, R., Morganti, R., \& Santantonio, L.\ 2005, \aap, 439, 487 




\bibitem[Drury(1983)]{D83} 
Drury, L. O'C. 1983, Rep. Prog. Phys., 46, 973


\bibitem[Dyer et al.(2001)]{DRB01} Dyer, K.~K., Reynolds, 
S.~P., Borkowski, K.~J., Allen, G.~E., \& Petre, R.\ 2001, \apj, 551, 439 



\bibitem[Ellison et al.(2001)]{ESG01} Ellison, D.~C., Slane, 
P., \& Gaensler, B.~M.\ 2001, \apj, 563, 191 






\bibitem[Elvis et al.(1994)]{E94} Elvis, M., et al.\ 1994, 
\apjs, 95, 1 






\bibitem[Fujita et al.(2007)]{FKY07}
Fujita, Y., Kohri, K., Yamazaki, R., \& Kino, M. 2007, \apj, 663, L61


\bibitem[Fukazawa et al.(2006)]{FBP06} Fukazawa, Y., 
Botoya-Nonesa, J.~G., Pu, J., Ohto, A., \& Kawano, N.\ 2006, \apj, 636, 698 



\bibitem[Fukazawa,  Makishima,  \& Ohashi(2004)]{FMO04} 
Fukazawa, Y., Makishima, K., \& Ohashi, T. 2004, PASJ, 56, 965

\bibitem[Godfrey et al.(2009)]{G09} Godfrey, L.~E.~H., et 
al.\ 2009, \apj, 695, 707 


\bibitem[Hardcastle et al.(2004)]{HHW04} Hardcastle, M.~J., 
Harris, D.~E., Worrall, D.~M., \& Birkinshaw, M.\ 2004, \apj, 612, 729 





\bibitem[Ito et al.(2008)]{IKK08} Ito, H., Kino, M., 
Kawakatu, N., Isobe, N., \& Yamada, S.\ 2008, \apj, 685, 828 				



\bibitem[Jiang et al.(2006)]{J06} Jiang, L., et al.\ 2006, 
\aj, 132, 2127 













\bibitem[Landau \&  Lifshitz(1959)]{LL59} 
Landau, L., \&  Lifshitz, F. M. 1959, Fluid Mechanics (London: Pergamon)





\bibitem[Manolakou et 
al.(2007)]{MHK07} Manolakou, K., Horns, D., \& Kirk, J.~G.\ 2007, \aap,
				474, 689 


\bibitem[Mathews 
\& Brighenti(2003)]{MB03} Mathews, W.~G., \& Brighenti, F.\ 2003, \araa, 41, 191 






\bibitem[Melrose(1980)]{M80} Melrose, D.~B.\ 1980, New 
York, Gordon and Breach Science Publishers, 1980.~430 p.,  


\bibitem[Moss  \& Shukurov(1996)]{MS96} 	
 Moss, D., \& Shukurov, A. 1996, \mnras, 279, 229

\bibitem[Mulchaey \& Zabludoff(1998)]{MZ98} 
Mulchaey, J. S., \& Zabludoff, A. I. 1998, ApJ, 496, 73


\bibitem[Ostorero et al.(2010)]{OMS10} Ostorero, L., et al.\ 
2010, \apj, 715, 1071 





\bibitem[Ostriker \& McKee(1988)]{OM88} Ostriker, J.~P., \& McKee, C.~F.\ 1988, Reviews of Modern Physics, 60, 1 












\bibitem[Reynolds et al.(2001)]{RHB01} Reynolds, C.~S., 
Heinz, S., \& Begelman, M.~C.\ 2001, \apjl, 549, L179 


\bibitem[Rybicki 
\& Lightman(1979)]{RL79} Rybicki, G.~B., \& Lightman, A.~P.\ 1979, New York, Wiley-Interscience, 1979.~393 p.,  








\bibitem[Schekochihin et al.(2005)]{SCK05} 	
 Schekochihin, A. A., Cowley, S. C., Kulsrud, R. M., Hammett, G. W.,
 \& Sharma, P. 2005, ApJ, 629, 139










\bibitem[Stage et al.(2006)]{SAH06} Stage, M.~D., Allen, 
G.~E., Houck, J.~C., \& Davis, J.~E.\ 2006, Nature Physics, 2, 614 


\bibitem[Stawarz et al.(2007)]{SCH07} Stawarz, {\L}., Cheung, 
C.~C., Harris, D.~E., \& Ostrowski, M.\ 2007, \apj, 662, 213 


\bibitem[Stawarz et al.(2008)]{SBM08} Stawarz, {\L}., 
Ostorero, L., Begelman, M.~C., Moderski, R., Kataoka, J., 
\& Wagner, S.\ 2008, \apj, 680, 911 


\bibitem[Tanaka et al.(2008)]{T08} Tanaka, T., et al.\ 
2008, \apj, 685, 988 


\bibitem[Vikhlinin,  Markevitch,  \& Murray(2001)]{VMM01} 	
 Vikhlinin, A., Markevitch, M., \& Murray, S. S. 2001, ApJ, 549, L47






\bibitem[Yamazaki et 
al.(2004)]{YYT04} Yamazaki, R., Yoshida, T., Terasawa, T., Bamba, A., \& Koyama, K.\ 2004, \aap, 416, 595 



\bibitem[Yaji et al.(2010)]{YTI10} Yaji, Y., Tashiro, M.~S., 
Isobe, N., Kino, M., Asada, K., Nagai, H., Koyama, S., 
\& Kusunose, M.\ 2010, \apj, 714, 37 











\end{thebibliography}
\end{document}